\documentclass[12pt,a4paper]{article}

\usepackage{hyperref}
\usepackage{amsmath}
\usepackage{amssymb}
\usepackage{mathrsfs}
\usepackage{bbm}
\usepackage{graphicx}
\usepackage{color}
\usepackage{enumerate}
\usepackage[normalem]{ulem}

\numberwithin{equation}{section}

\begin{document}

\title{Holography without strings?}

\author{
  Donald Marolf\\
  \\
  {\it Department of Physics, University of California at Santa Barbara,}\\
    {\it Santa Barbara, CA 93106, U.S.A.} \\
  }

\date{\today}

\maketitle

\begin{abstract}

A defining feature of holographic dualities is that, along with the bulk equations of motion, boundary correlators at any given time $t$ determine those of observables deep in the bulk.  We argue that this property emerges from the bulk gravitational Gauss law together with bulk quantum entanglement as embodied in the Reeh-Schlieder theorem.  Stringy bulk degrees of freedom are not required and play little role even when they exist.   As an example we study a toy model whose matter sector is a free scalar field.  The energy density $\rho$ sources what we call a pseudo-Newtonian potential $\Phi$ through Poisson's equation on each constant time surface, but there is no back-reaction of $\Phi$ on the matter.  We show the Hamiltonian to be essentially self-adjoint on the domain generated from the vacuum by acting with boundary observables localized in an arbitrarily small neighborhood of the chosen time $t$.  Since the Gauss law represents the Hamiltonian as a boundary term, the model is holographic in the sense stated above.
\end{abstract}

 \maketitle

\newpage

\tableofcontents



\section{Introduction}
\label{intro}

Holographic dualities  \cite{Banks:1996vh,Maldacena:1997re} are settings where one theory (the bulk) is dual to a second theory (the dual field theory, or DFT) living on a lower dimensional spacetime.  In simple cases the DFT spacetime can be identified with the boundary of the bulk, and local DFT operators can be identified  \cite{Witten:1998qj} with boundary limits of bulk operators.   For this to be a true duality any bulk operator must, at least in principle, be expressible in terms of DFT operators. This suggests that all bulk operators are in fact determined by their boundary values.

While this property may sound striking at first, in a free bulk quantum field theory it is actually straightforward to show that all bulk operators can be written in terms of their boundary values.  The point is that signals in the bulk eventually travel outward and reach the boundary.\footnote{At least for signals outside black holes. See \cite{Freivogel:2005qh,Marolf:2008tx,Czech:2012be,Marolf:2012xe,VanRaamsdonk:2013sza} for comments on possible limitations of the duality related to eternal black holes.  Inside a black hole that forms from collapse, one may evolve signals backward in time until they reach the boundary, though the resulting perturbation theory becomes difficult to control at times long after the collapse \cite{Susskind:2012uw,Almheiri:2013hfa}.}  The free result may then be corrected perturbatively for bulk interactions which, in familiar examples, corresponds to performing a $1/N$ expansion in the DFT \cite{Banks:1998dd,Balasubramanian:1998de,Bena:1999jv,Hamilton:2005ju,Kabat:2011rz,Heemskerk:2012mn,Kabat:2012av}.

The interesting point about holographic theories is that they take this observation one step further.   Since the DFT is a self-contained theory which evolves deterministically under its own Hamiltonian $H_{DFT}$, for any DFT operator ${\cal O}_{DFT}$ at any time $t_1$ we may write
\begin{equation}
{\cal O}_{DFT}(t_1) = e^{iH_{DFT}(t_1-t)} {\cal O}_{DFT}(t)
^{-iH_{DFT}(t_1-t)},
\end{equation}
expressing ${\cal O}_{DFT}(t_1)$
in terms of DFT operators at time $t$.  Combining this with the above reasoning suggests that any bulk observable (defined, say, at some time $t$) can in fact be written in terms of boundary limits of bulk operators at the {\it same} time $t$. We call this strengthened conjecture holography of information and note that it is closely related to discussions of so-called precursors in \cite{Polchinski:1999yd,Susskind:1999ey}.   We also use the term boundary observable for the relevant boundary limits of bulk operators; i.e., for what were called extrapolated observables in \cite{Harlow:2011ke}.  In contrast, observables in the dual field theory will be called DFT observables.

Holography of information would be manifestly false in any local non-gravitational quantum field theory.  Since all equal time commutators between bulk and boundary must vanish due to the spacelike separation of the operators involved, no non-trivial bulk commutator could be reproduced by boundary observables.   And even classical gravity allows locally specified initial data, so that the full classical solution is certainly not determined by its boundary values at the single time $t$.

Yet there is a sense in which holography of information becomes natural for gravitational systems at the quantum level \cite{Marolf:2008mf}.  It is well-known that the on-shell Hamiltonian $H$ of classical gravity is a pure boundary term \cite{Deser:1960zzc,Deser:1961zza,Arnowitt:1962hi,Regge:1974zd,Abbott:1981ff}, a fact that is deeply related to having a non-trivial notion of diffeomorphism invariance (see e.g. \cite{Andrade:2010hx}).  This is essentially the relativistic version of the gravitational Gauss law.  If $H$ remains a boundary term in the quantum theory, then the final step in the above argument can be made directly in the bulk:  since $H$ itself is a boundary observable, writing ${\cal O}(t_1) = e^{iH(t_1-t)} {\cal O}(t)
^{-iH(t_1-t)}$ immediately expresses the boundary observable ${\cal O}$ at any time $t_1$ in terms of boundary observables at time $t$ -- a property called boundary unitarity in \cite{Marolf:2008mf}\footnote{In perturbation theory, one thinks of the gravitational Gauss law as simply dressing each particle state with an appropriate gravitational Coulomb field. One may ask how this leads to boundary unitarity.  The point seems to be that the Coulomb tails are rich enough to allow states that can be distinguished by boundary observables at any time $t_1$ to also be distinguished by boundary observables at time $t$.  The total energy plays a role in this process, as do boundary observables whose commutators with $H$ are non-zero.  The basic mechanism for this was discussed in \cite{Balasubramanian:2006iw}. }.    For simplicity we have used notation appropriate to time-independent Hamiltonians, though making the exponentials path-ordered generalizes the argument to the time-dependent case as well.

It is notable that this argument makes no mention of any stringy degrees of freedom\footnote{We use this term to also refer to the towers of fields in higher spin gravity \cite{Fradkin:1987ks,Vasiliev:1999ba,Klebanov:2002ja}.}.  This stands in contrast both to the central role of strings in implementing details of the known gauge/gravity correspondences and to past suggestions of their relevance to this issue \cite{Polchinski:1999yd,Giddings:2001pt,Freivogel:2002ex}.  On the other hand, our argument for information holography may appear rather formal, and the fact that exponentiating the Hamiltonian is equivalent to solving the equations of motion may suggest that further subtleties remain to be considered.  For example,  although the on-shell Hamiltonian $H$ can be written as an integral over boundary limits of the gravitational field at any time $t$ and is thus clearly a boundary observable at this time, it is natural to ask whether the exponentiated operator $e^{iHt}$ also qualifies as a boundary observable at the single time $t$.  The goal of our work below is to carefully argue that it does, and that stringy degrees of freedom continue to play no significant role.

This conclusion may come as a surprise to some readers.   A counter-argument that strings should be instead critical may be given as follows.   When the DFT is a gauge theory, completeness of Wilson loops in the DFT observable algebra suggests that bulk observables can in fact be written in the DFT as a sum over (perhaps suitably decorated) spacelike Wilson loops at any given time $t$.  Transcribing this result to the bulk would then involve strings \cite{Maldacena:1998im}. But the problem with this supposed example is two-fold.  First, the sum over Wilson loops is also rather formal, involving Wilson loops with large amounts of structure on arbitrarily small scales (see comments in section 4.4.3 of \cite{Heemskerk:2012mn}).  Second, it is unclear what special role such strings could play in the bulk.  Indeed, in the limit of weak string coupling one would expect to be able to use some form of string field theory to treat strings in much the same way as very heavy particles.  But, as noted above, in standard quantum field theory information is non-holographic no matter how heavy the particles may happen to be.  So while local boundary operators certainly mix with more stringy operators when the DFT is a gauge theory, the fundamental significance of this effect remains unclear.

Let us therefore return to the bulk Hamiltonian and ask to what extent $e^{iHt}$ might contain information beyond that in $H$ itself.  This rather technical sounding question clearly demands a technical answer.  We thus adopt the mathematical physics point of view and note that, since $H$ is unbounded, it can be defined only on some dense domain ${\cal D}$ smaller than the full Hilbert space ${\cal H}$ of the quantum theory.  In contrast,  $e^{iHt}$ is bounded and thus continuous, and defined on all of ${\cal H}$.  So long as $H$ is truly self-adjoint (see e.g. \cite{Reed:1980}), this difference is usually considered to be trivial.  The operator $H$ then has a complete set of orthogonal eigenstates on which $e^{iHt}$ is both defined and bounded, so the linear extension of ${\cal D}$ to ${\cal H}$ is unique.  Similarly, in constructing Hamiltonians one often first defines $H$ on a dense domain ${\cal D}$ that, while it might not in itself make $H$ self-adjoint, may still be large enough to guarantee that $H$ has a unique self-adjoint extension.  $H$ is then said to be essentially self-adjoint on ${\cal D}$ and the operator $e^{iHt}$ is again uniquely determined.

We ask below whether, given some reasonable choice of bulk quantum state $|\psi \rangle$, the domain ${\cal D}$ defined by acting on $|\psi \rangle$ with products of local boundary observables\footnote{Perhaps in stringy examples supplemented by less-local boundary observables associated with large strings dual to Wilson loops in the DFT.} at some given time $t$ is one that makes $H$ essentially self-adjoint.  We study the case where $|\psi \rangle$ is the ground state $|0 \rangle$ of the bulk theory.  Indeed, one might expect that the DFT Hamiltonian is essentially self-adjoint on the domain defined by applying all local DFT operators (perhaps supplemented in a gauge theory by spacelike Wilson loops) at time $t$ to the DFT vacuum, so we ask whether a bulk dual of this property can be identified\footnote{This interesting question was brought to our attention by Bob Wald.}.

We show below that this is so.  Our argument uses properties related to the Reeh-Schlieder theorem which states (see e.g. \cite{Haag:1992hx}) that by acting on $|0\rangle$, or in fact any state of bounded energy, with operators localized in any arbitrarily small region of spacetime one may approximate any state in a quantum field theory Hilbert space to arbitrary precision.  This highly quantum property has no classical counterpart and stems from the strong entanglements required to keep the energy small. Section \ref{ex} studies the simple toy model of a free scalar field $\chi$ in Anti-de Sitter (AdS) space, supplemented by an extra fully constrained field, the pseudo-Newtonian potential $\Phi$, which allows us to treat the Hamiltonian as a boundary observable \cite{Andrade:2010hx}.  The introduction of this $\Phi$ models the gravitational Gauss law of true holographic systems but plays no further role in our analysis.   We show that $H$ is essentially self-adjoint on the domain defined by acting  in any neighborhood of $t=0$ with boundary limits of $\chi$ on the vacuum $|0\rangle$.  We then interpret this result and discuss generalizations in section \ref{disc}.

\section{Holography in pseudo-Newtonian gravity}

\label{ex}

As mentioned above, we build our model from a free scalar field $\chi$ of mass $m$ on the $(d+1)$-dimensional (global) Anti-de Sitter (AdS) space $AdS_{d+1}$  with metric
\begin{equation}
ds^2 = -(1+r^2/\ell^2) dt^2 + \frac{dr^2}{1+r^2/\ell^2} + r^2 d\Omega^2.
\end{equation}
We use $\Omega$ to denote a point on $S^{d-1}$ with $d\Omega^2$ the associated unit-radius round metric.  Due to the timelike conformal boundary of AdS, we must impose boundary conditions to make the dynamics well-defined.  We will use AdS-invariant boundary conditions associated with choosing a so-called\footnote{This terminology is useful and familiar to practitioners of AdS/CFT, though we emphasize that we study only the bulk.  At no point do we assume the existence of any dual CFT.} conformal dimension $\Delta$ satisfying \cite{Breitenlohner:1982jf,Breitenlohner:1982bm,Mezincescu:1984ev}
$\Delta = -\frac{d}{2} \pm \sqrt{\left(\frac{d}{2} \right)^2 + m^2}$.
We require our theory to be ghost-free, which imposes $\Delta > \frac{d-2}{2}$ and makes the choice of boundary conditions unique for $m^2 \ge -\frac{d^2}{4} +1$ (see \cite{Andrade:2011dg} which builds on \cite{Breitenlohner:1982jf,Breitenlohner:1982bm,Balasubramanian:1998sn,Klebanov:1999tb,Ishibashi:2004wx,Compere:2008us}).

The quantum field $\chi$ may be written
\begin{equation}
\chi(x,t) = \sum_{n, \vec \ell} \left( a^\dagger_{n, \vec \ell} u^*_{n, \vec \ell} + a_{n, \vec \ell} u_{n, \vec \ell}\right) \ \ \ {\rm for} \ \ \
u_{n, \vec \ell}(t, \Omega, r) = \frac{e^{-i\omega t}}{\sqrt{\omega}} Y_{\vec \ell}(\Omega) R_{n, \vec \ell} (r),
\end{equation}
where $a^\dagger_{n, \vec \ell}, a_{n, \vec \ell}$ are the usual creation/annihilation operators, $n \ge 0$ is an integer, $\vec \ell$ labels spherical harmonics $Y_{\vec \ell}(\Omega)$ on $S^{d-1}$ with principle quantum number $|\vec \ell|$,
$\omega = (2\Delta + |\vec \ell| + 2n)$, and the $u_{n, \vec \ell}$ are a complete set of Klein-Gordon-normalized positive-frequency wavefunctions.
We use $x,y$ to denote the $d$ spatial coordinates and will always separately indicate dependence on the time $t$.
The radial profiles $R_{n, \vec \ell}$ are hypergeometric \cite{Breitenlohner:1982jf,Breitenlohner:1982bm,Balasubramanian:1998sn}, though the only properties we require are that $R_{n, \vec \ell} = O(r^{-\Delta})$ at large $r$ and that $\beta_{n, \vec \ell}: = \lim_{r\rightarrow \infty} r^\Delta R_{n, \vec \ell}$ scales like $n^\Delta$ at large $n$ for fixed $\vec \ell$.
Thus the natural boundary operator $X(\Omega, t) := \lim_{r \rightarrow \infty} r^\Delta \chi(r, \Omega, t)$  takes the form
\begin{equation}
\label{Xop}
X(\Omega, t) = \sum_{n, \vec \ell} \left( a^\dagger_{n, \ell} U^*_{n, \vec \ell} + a_{n, \ell} U_{n, \vec \ell} \right), \  \ \ {\rm for} \ \ \
U_{n, \vec \ell}(\Omega,t) = \frac{\beta_{n, \vec \ell}}{\sqrt{\omega}} e^{-i\omega t} Y_{\vec \ell}(\Omega),
\end{equation}
where the coefficient of $Y_{\vec \ell}$ scales like $n^{\Delta - 1/2}$ at large $n$.

To complete our model, we also introduce the non-local functional
\begin{equation}
\label{Green}
\Phi(x,t) = \int d^dx \ G(x,y) \rho(y,t)
\end{equation}
of $\chi$, where $\rho= \frac{1}{2(1+ r^2/\ell^2)} \dot \chi^2 + \frac{1}{2} g^{ij} \partial_i \chi \partial^i \chi + \frac{m^2}{2} \chi^2 $ is the (scalar) energy density of our system and $G(x,y)$ is the Dirichlet Green's function for Poisson's equation
\begin{equation}
\label{Poisson}
D_i D^i \Phi = - \rho.
\end{equation}
Here $D$ is the covariant derivative on each constant $t$ surface which we label by coordinates $x^i$ for $i=1 \dots d$.   The boundary conditions on $\chi$ ensure that $\rho$ decays fast enough for \eqref{Green} to converge.  Thus \eqref{Green} is equivalent to requiring $\Phi$ to satisfy \eqref{Poisson}
on each constant $t$ surface with boundary condition $\Phi = 0$ at $r=\infty$.  Although the definition \eqref{Green} may look rather contrived in our model, equation \eqref{Poisson} models the gravitational Gauss law of known holographic theories.

Due to the formal similarity of \eqref{Poisson} to Newtonian gravity, we call $\Phi$ the pseudo-Newtonian potential.  Differences from actual Newtonian gravity include that we take $\Phi$ to have no effect on the matter fields, that our matter theory is relativistic, and that the background geometry is AdS.  Of more importance however is the fact that our $\rho$ is the full energy density of the matter theory, and not just some non-relativistic notion of mass-density.  As noted in \cite{Andrade:2010hx}, Gauss' law then implies that the Hamiltonian may be written as a boundary term:
\begin{equation}
\label{Hbound}
H = - \lim_{R \rightarrow \infty} \int_{r = R, t = const} \hat r^i \partial_i \Phi \ R^{d-1} d \Omega,
\end{equation}
where $\hat r^i$ is the outward-pointing unit normal at the boundary and $d\Omega$ is the volume element on the unit $(d-1)$-sphere.  While this statement is not particularly deep for our toy system, it models a central property of metric theories of gravity which distinguish them from systems with more trivial notions of diffeomorphism invariance \cite{Andrade:2010hx}. Note that \eqref{Hbound} generates time-translations of both $\chi$ and $\Phi$.

For any time interval $[-\epsilon,\epsilon]$ no matter how small,
we wish to show essential self-adjointness of the operator $H_{\cal D}$ given by restricting the true self-adjoint Hamiltonian $H$ to the domain ${\cal D}$ constructed by acting on the vacuum $|0\rangle$ with boundary observables $X(\Omega, t)$ smeared with test functions supported in $[-\epsilon, +\epsilon]$.  There is also a corresponding boundary observable defined by $\Phi$, though the $X(\Omega, t)$ turn out to be sufficient for our argument.  The term essentially self-adjoint means that $H$ is in fact the {\it unique} self-adjoint operator on the full Hilbert space ${\cal H}$ whose restriction to ${\cal D}$ gives $H_{\cal D}$.  Our test functions will be sufficiently differentiable that states in ${\cal D}$ do indeed lie in the domain of $H$, and also in the domain of $H^n$ for any finite $n$.

Since the spectrum of $H$ is discrete, it has a complete set of normalized eigenstates $|E\rangle$ with eigenvalues $E$.  Let us suppose that (as we will shortly establish) for every $|E\rangle$ there is a sequence of approximations $|\psi_N(E)\rangle \in {\cal D}$ such that as $N \rightarrow \infty$ we have
\begin{enumerate}[i)]
\item{} $|\psi_N(E) \rangle \rightarrow |E\rangle$ in the Hilbert space norm and
\item{} $\langle \psi_{N}(E) |(H-E)^2 | \psi_{N}(E) \rangle \rightarrow 0$.
\end{enumerate}
Since $ | \psi_{N}(E) \rangle \in {\cal D}$, requirement (ii) implies that the same limit vanishes when $H$ is replaced by $H_{\cal D}$, and thus by any extension $\tilde H$ of $H_{\cal D}$.  But if $\tilde H$ is self-adjoint, it also has a complete set of eigenvectors $|\tilde E \rangle$ with eigenvalues $\tilde E$. For simplicity of notation we assume that $\tilde H$ has again a purely discrete spectrum, though any continuous components are readily included through the usual replacements of sums by integrals.  We may then expand our approximations $|\psi_N(E) \rangle$ in terms of the $|\tilde E \rangle$ as  $|\psi_N(E) \rangle = \sum_{\tilde E} \psi_N(E,\tilde E) |\tilde E \rangle$.   Replacing $H$ by $\tilde H$ in (ii) shows that
\begin{equation}
\sum_{\tilde E} (\tilde E - E)^2 |\psi_N(\tilde E, E)|^2 \rightarrow 0 \ \ {\rm as} \ \ N \rightarrow \infty.
\end{equation}
Since the terms in this sum are all positive, each one must in fact vanish separately.  On the other hand, requirement (i) and normalizeability of $|\tilde E\rangle$ imply that $\psi_N(E,\tilde E) \rightarrow \langle \tilde E | E\rangle$.  It follows that $(\tilde E -E)^2 |\langle \tilde E | E \rangle|^2 =0$ for all $\tilde E,E$.  We conclude that each $E$ is one of the eigenvalues $\tilde E$ with $|E \rangle = |\tilde E \rangle$ and that $H =  \tilde H$.

Thus we need only show properties (i) and (ii) above. As a first step, consider the frequency-space function $\tilde f_0(\omega) = \frac{2}{\omega}\sin(\omega \epsilon)$.
With appropriate conventions this is the Fourier transform of $f_0(t) = [\theta(t+\epsilon) - \theta (t-\epsilon)]$, the characteristic function of our interval. For positive integers $N$ we may recursively define the related frequency-space functions $\tilde f_N := \frac{\tilde f_{N-1}}{1 +  \frac{\omega \epsilon}{\pi N}}$.   That these are the Fourier transforms of $C^{N-1}$ functions $f_N$ supported on $[-\epsilon,\epsilon]$ follows either from the $C^{N-1}$ version of the Paley-Wiener theorem or by noting that, up to an overall factor of $\frac{\epsilon}{\pi N}$, the function $f_N$ is generated from $f_{N-1}$ by the following three-step procedure: First multiply by $e^{ i\frac{\pi N}{\epsilon}t}$, which enacts a translation in Fourier space, then integrate the result from $-\infty$ to $t$ (equivalent to dividing by $\omega$ in Fourier space), and finally multiply by  $e^{- i\frac{\pi N}{\epsilon} t}$.  Because $\tilde f_{N-1}$ vanishes at $\omega = - \pi N/\epsilon$, the integral in the second step vanishes in the limit $t \rightarrow \infty$. Since $f_{N-1}$ is supported on $[-\epsilon,\epsilon]$, the integral in fact vanishes for $|t| > |\epsilon|$ so that $f_N$ is supported on the same interval.

Note that $\tilde f_N$ is $1$ at $\omega = 0$ for all $N \ge  0$.  In contrast, for any $\omega > 0$ the value $\tilde f_N(\omega)$ is a strictly decreasing function of $N$ which vanishes in the limit $N \rightarrow \infty$. This vanishing may be seen from the fact that $\ln \prod_{i=1^N} (1 + \frac{\omega \epsilon }{\pi N}) \sim \int  dk \  \ln (1 + \frac
{ \omega \epsilon}{\pi k}) \sim \int dk \ \frac{\omega \epsilon }{\pi k} \sim \ln k$, where $\sim$ indicates similar behavior at large $N, k$.

We are now ready to demonstrate (i) and (ii) for the 1-particle state $|0, \vec \ell \rangle$ of our theory which minimizes the 1-particle energy for fixed angular momentum $\vec \ell$.  We take this state to have frequency $\bar \omega$ and note that, because $\bar \omega$ is the lowest relevant frequency,  $g_N(t) = e^{i \bar \omega t} f_N(t)$ gives a sequence of smooth $C^{N-1}$ functions supported on $[-\epsilon,\epsilon]$ which approximate the Kronecker delta function $\delta_{\omega, \bar \omega}$ when evaluated on 1-particle states with angular momentum $\vec \ell$.  In particular, the states
\begin{equation}
\label{approx}
|\psi_N \rangle := \int dt d\Omega \ g^*_N(t) Y^*_{\vec \ell} (\Omega) X(t, \Omega) | 0 \rangle
\end{equation}
satisfy properties (i) and (ii) for $|E\rangle = |0, \vec \ell \rangle$.  The scaling mentioned below \eqref{Xop} implies that the action of $H$ on such states is well-defined for all $N > 2\Delta +1$, and that higher powers of $H$ become well-defined at correspondingly higher $N$.  If desired, one could deform the $g_N(t)$ into smooth functions supported on $[-\epsilon,\epsilon]$ which again satisfy (i) and (ii) but for which $H^n$ is well-defined on $|\psi_N \rangle $ for all $n, N$.

The $k$th 1-particle state with angular momentum $\vec \ell$ may now be iteratively approximated by a similar method. We need only define functions $g_{k,N}$ in analogy to $g_N(t)$ by setting $\bar \omega$ equal to the $k$th frequency and taking care to subtract off (to good approximation) components proportional to lower one-particle states studied previously\footnote{An alternate approach is to define $\tilde g_{k,N}(\omega) = F_k(\omega)e^{-i\bar \omega} \tilde f_N(\omega)$ where $F_k$ is a fixed polynomial in $\omega$ that takes the value $1$ at the $k$ frequency $\bar \omega$ and vanishes at each lower 1-particle frequency.}.  Since arbitrarily good approximations (in norm) to the lower states have already been constructed, the effect of this subtraction on the higher states can be made negligible.   States with $n$-particles can be constructed by applying the operator in \eqref{approx} $n$-times, again taking care to subtract off appropriate approximations to any $m$-particles states with $m < n$ produced in this way.  Putting our results together, we conclude that $H$ is the unique self-adjoint extension of $H_{\cal D}$ to the full Hilbert space.

\section{Discussion}
\label{disc}

We argued above that, for linear scalar fields on AdS$_{d+1}$, boundary operators in any neighborhood of $t=0$ act on the vacuum $|0\rangle$ to define a domain ${\cal D}$ on which the Hamiltonian $H$ is essentially self-adjoint.
The particular argument involved constructing good approximations to bulk energy eigenstates. This construction may also be relevant to the particular issues discussed in \cite{Giddings:1999jq,Gary:2009ae,Gary:2009mi,Gary:2011kk}.

We supplemented this model with a pseudo-Newtonian potential $\Phi$ whose Gauss law promotes the Hamiltonian $H$ to an additional boundary observable.  This gives a class of examples where the action of the exponentiated operators $e^{iHt}$ are uniquely defined by correlators of boundary observables arbitrarily close to $t=0$, and thus where such correlators determine all bulk correlation functions -- even when the arguments are localized far from the AdS boundary.  Information in our models is then holographic in the sense used here.

In addition to the manifest use of the gravitational Gauss law, the key feature in our argument was the strong vacuum entanglement required by quantum field theory.  This is what makes $H$ essentially self-adjoint on ${\cal D}$.  As in the Reeh-Schlieder theorem, the same result would follow using any state of bounded energy.\footnote{However, again as for Reeh-Schlieder, one can also find less-entangled states where acting with $X$ (or even Wick powers thereof) is no longer sufficient to generate a domain of the desired form.  For example, if $f(x,t)$ is a smooth test function whose support is spacelike separated from the relevant region of the boundary, such a state is given by projecting $|0\rangle$ onto some range of eigenvalues of ${\cal O} = \int \phi (x,t) f(x,t) d^{d}xdt$.  Since ${\cal O}$ commutes with the relevant smeared operators $X$, they cannot change the eigenvalues of ${\cal O}$ and the domain generated is not even dense in ${\cal H}$.  On the other hand, a larger domain can clearly be formed by also acting with $H$, or more generally with $\Phi(\Omega, t)$, which provide additional boundary observables whose commutator with ${\cal O}$ is non-zero.  This observation suffices to remove direct tension with completeness of the local DFT observables, though whether the resulting domain actually makes $H$ essentially self-adjoint remains an interesting question for future study.}

It is worth emphasizing that the required entanglement is a highly quantum phenomenon, with no analogue in classical physics.  In particular, given any density function on a classical field theory phase space, the space of densities generated by multiplying by functions of the boundary data is far from complete\footnote{At the mathematical level, a closer analogy would be to ask about the space of densities generated by taking Poisson brackets with boundary observables as well as a acting by multiplication.  To get some insight into this question, we may consider the Harmonic oscillator. Since the key to our argument above was the ability to discard negative energy parts of operators, the fact that no real phase space function can have vanishing Poisson bracket with $a = \frac{x}{\sqrt{m\omega}}-i\omega p$ suggests that there is again no classical analogue of our property.  In any case, the mathematical question raised in this footnote is physically quite different than that studied in Reeh-Schlieder as taking Poisson brackets does not generally preserve positivity of the phase space density.}.

Our discussion also highlights the difference between acting with general operators and coupling to boundary sources.  Since the latter enact unitary transformations, the set of states generated from the vacuum by adding sources for our boundary field $X$ (or any of its Wick powers) in a small time interval $(t_1,t_2)$ is not dense.  In particular, given the operator ${\cal O}$ just defined and localized at spacelike separations from the relevant piece of the boundary, this action cannot change the expectation value of e.g. $B= \exp(i\lambda {\cal O})$ (or any other bounded function of ${\cal O}$).  Thus one cannot approximate eigenstates of $B$ whose eigenvalues differ from $\langle 0 | B| 0 \rangle$.

Further study of similar pseudo-Newtonian models may shed additional light on information holography.  We focused here on global AdS backgrounds both for concreteness and to provide a closer connection to gauge/gravity duality.  But the argument applies much more generally.  Indeed, the only features we used were discreteness of the spectrum of frequencies $\omega$ (though the continuous spectrum case is similar) and, more importantly, that the coefficients in the expansion \eqref{Xop} of the natural boundary operators $X$ in terms of creation/annihilation operators grew no faster than some fixed power law at large $\omega$.  Both properties hold for linear quantum fields in bounded regions of Minkowski space, or in fact in any other example (whether lattice or continuum) with a finite-distance boundary at finite redshift.  It would be interesting to extend our results to interacting theories, perhaps treated perturbatively.

While potentially valuable as toy models of information holography, we remind the reader that pseudo-Newtonian systems are also very different from standard gauge/gravity dualities. Perhaps most importantly, the analogue of the DFT is not only non-relativistic but in fact highly non-local.  The point here is that the DFT energy density operator is $\varepsilon_{DFT} = - \lim_{r \rightarrow \infty} r^{d-1} \hat r^i \partial_i \Phi$, and since $\Phi(x,t) = \int G(x,y) \rho(y,t)$ is an integral over all space commutators $[\varepsilon(\Omega,t), \varepsilon(\Omega', t)]$ will not vanish even at equal times.  Indeed, it was precisely to make this point that such models were mentioned in \cite{Andrade:2010hx}.

We have stressed that our model contains no stringy degrees of freedom.  We conclude that stringy dynamics is not required for information holography and suggest that, even when they are present in holographic theories, such degrees of freedom may play little direct role. It was of course already known that other critical properties of the DFT are not directly connected to strings.  These include the vanishing of commutators at spacelike separation (i.e., locality, which follows from a quantum version of \cite{Gao:2000ga}) and the existence in appropriate cases of a DFT stress tensor \cite{Witten:1998qj}.  Thus our work here strengthens the argument that any UV complete theory of gravity will be holographic, even if it contains no strings.\footnote{This idea has certainly been mentioned many times.  The author would welcome suggestions as to appropriate references.}

On the other hand, the existence of bulk strings is intimately related to the gauge theoretic nature of the DFTs that arise in known examples \cite{Maldacena:1998im}.  And since the notion of information holography studied here is rather abstract, it may be that such gauge theoretic DFTs somehow yield a simpler connection in which strings play a role.  If so, it would be useful to understand this in detail.  It is similarly plausible that stringy bulk physics is required for the DFT to admit a Lagrangian formulation, or perhaps for the theory to be `simple' in some more general sense that would include examples like the $d=6$ (2,0) superconformal theory of \cite{Berkooz:1997cq}.

We have focussed on the property we call information holography.  But we remind the reader that another even more striking property of familiar gauge/gravity dualities \cite{Banks:1996vh,Maldacena:1997re} is their finite density of states dual to bulk black holes.  This feature remains a complete mystery from the bulk point of view.  As has been noted by many authors, the fundamental appearance of the Planck scale in the Bekenstein-Hawking entropy again suggests a gravitational mechanism having little to do with strings.  Identifying this mechanism is a worthy goal that may require models quite different from those considered here.

\section*{Acknowledgements}
This paper is a direct result of repeated interaction with Bob Wald.  I also thank Aron Wall for related discussions and Gary Horowitz, Joe Polchinski, Mark Srednicki, and Aron Wall for comments on an early draft. This work was supported in part by the National Science Foundation under Grant No PHY11-25915 and by funds from the University of California.

\providecommand{\href}[2]{#2}\begingroup\raggedright\endgroup

\end{document}